# Prognosis of COVID-19 using Artificial Intelligence: A Systematic Review and Meta-analysis

SaeedReza Motamedian[1,2], Sadra Mohaghegh[1,2*], Elham Babadi Oregani[2], Mahrsa Amjadi[2], Parnian Shobeiri1, Negin Cheraghi[2], Niusha Solouki[2], Nikoo Ahmadi[2], Hossein Mohammad-Rahimi[1], Yassine Bouchareb[3], Arman Rahmim[4,5]

[1]Topic Group Dental Diagnostics and Digital Dentistry, ITU/WHO Focus Group AI on Health, Berlin, Germany

[2]Dental Research Center, Research Institute of Dental Sciences, Shahid Beheshti University of Medical Sciences, Tehran, Iran

[3]Sultan Qaboos University, College of Medicine and Health Sciences, Radiology and Molecular Imaging, Muscat, PO Box 35, PC 123, Oman

[4]Department of Radiology, University of British Columbia, Vancouver, BC, Canada

[5]Department of Physics, University of British Columbia, Vancouver, BC, Canada

**Corresponding Author**: Sadra Mohaghegh, DDS, Tehran, Iran, mohaghegh.sa77@gmail.com, +989120332633

# Abstract

**Purpose:** Artificial intelligence (AI) techniques have been extensively utilized for diagnosing and prognosis of several diseases in recent years. This study identifies, appraises and synthesizes published studies on the use of AI for the prognosis of COVID-19.

**Method:** Electronic search was performed using Medline, Google Scholar, Scopus, Embase, Cochrane and ProQuest. Studies that examined machine learning or deep learning methods to determine the prognosis of COVID-19 using CT or chest X-ray images were included. Polled sensitivity, specificity area under the curve and diagnostic odds ratio were calculated.

**Result:** A total of 36 articles were included; various prognosis-related issues, including disease severity, mechanical ventilation or admission to the intensive care unit and mortality, were investigated. Several AI models and architectures were employed, such as the Siamense model, support vector machine, Random Forest , eXtreme Gradient Boosting, and convolutional neural networks. The models achieved 71%, 88% and 67% sensitivity for mortality, severity assessment and need for ventilation, respectively. The specificity of 69%, 89% and 89% were reported for the aforementioned variables.

**Conclusion:** Based on the included articles, machine learning and deep learning methods used for the prognosis of COVID-19 patients using radiomic features from CT or CXR images can help clinicians manage patients and allocate resources more effectively. These studies also demonstrate that combining patient demographic, clinical data, laboratory tests and radiomic features improves models' performance.

# 1 Introduction

COVID-19 began in early December 2019 and spread rapidly worldwide (1). The pandemic caused significant shortcomings, abrasion and burnout in primary and tertiary care healthcare institutions (2). The increase in hospital admissions has led to a remarkable increase in human errors (3, 4). Consequently, the care needed for many patients during peak periods could not be adequately provided. Rapid diagnosis of COVID-19 and determination of the severity of infection enable healthcare professionals to better control the virus spread and manage increased hospital overloads, aiming to improve the quality of treatments (5). Despite the recent ease in the COVID-19 situation, the lessons learned will help better manage future pandemics.

Triage is essential in patient management, alleviating the pressure on medical departments (6). COVID-19 patients indicate various presentations and outcomes, ranging from asymptomatic to critical situations that may lead to death (7). Based on the severity of the infection, it is essential to determine whether patients can receive care at home or should be admitted to COVID wards or intensive care units (ICU). It is also important to diagnose patients who require mechanical ventilation (whether non-invasive or via intubation) (8). Accordingly, prediction models can help triage systems by automatically combining predictors to estimate the severity, ventilation or intensive care needed and the possibility of death, hence allocating adequate resources (5). Indeed, determining these factors at the early stage helps clinicians prioritize patients during peak periods (9).

Several studies have already used prediction methods (e.g., rule-based scoring systems or advanced machine learning models) to accelerate patient assessment and ease pressure on frontline departments (10, 11). Conventionally, radiomic features are extracted from the previously segmented region of interest (11). Using deep learning models, features can be implicitly derived from images without the necessity of defining a region of interest. Quantitative features extracted from images can help identify relevant disease biomarkers, impact the clinical decision-making process, and provide means of predicting lesions' growth and characteristics (12).

Artificial intelligence (AI) is the science of making intelligent programs or applications that mimic human intelligence, perform the rapid assessment, and make accurate decisions (13). AI can analyze a large amount of data in a short time and potentially provides accurate outcomes (14). Recent deployment of AI models can be justified by considering its merits. AI can alleviate the need of doing some repetitive tasks by physicians and technical staff, accelerate time-consuming processes, enhance quantification and interpretation, improve diagnostic reproducibility, and provide clinically relevant information (15). Accordingly, AI techniques have been widely used for clinical purposes such as diagnosis, analysis of medical images, extensive data collection, research and clinical trials, management of intelligent health records and prediction of outbreaks (16-19). On the other hand, concerns about the reproducibility, generalizability, and explainability of AI models remain to be solved, presently hindering AI translation and implementation in clinical practice (20, 21).

Several studies examined the performance of these models in prognosis using computed tomography (CT) or chest X-ray (CXR) images (22-25). Systematic reviews on the application of AI for screening or diagnosis, prevention and treatment planning of COVID-19 have been performed recently (26-29). Unlike these reviews, this systematic review focuses on the use of AI for the prognosis of COVID-19 and quantitatively analyzes the performance of the deployed models using variables such as sensitivity, specificity and area under the curve (AUC) (30). Building on the existing knowledge from several recent studies on this topic, we attempted to provide an updated review with a special focus on prognosis, infection severity, need for ventilation or ICU and mortality, and also report on the most commonly used performance parameters, including AUC, accuracy, sensitivity and specificity.

# 2 Methods

## 2.1 Protocol and Registration

The question of this study, according to PICO format, was as follows: To compare the function (O) of AI models (I) in determining the prognosis of COVID-19 patients (P) with the specified ground truth (C). The study was carried out according to the preferred reporting items for systematic review and meta-analysis of diagnostic test accuracy studies (PRISMA-DTA) guidelines (31). The review was registered in Prospero with the number of CRD42022351594.

## 2.2 Eligibility Criteria

The inclusion and exclusion criteria used for selecting the articles are presented in Table 1.

## 2.3 Information Sources

The electronic search was performed in PubMed, Google Scholar, Scopus, Embase, Web of Science, Cochrane and ProQuest databases for English articles published before March 2023.

## 2.4 Search

The queries are indicated in Table 2. English articles were included, and no restriction was set on the publication date. Also, no filter was used for the type of study.

## 2.5 Study Selection

The electronic search results were entered into Endnote 20 software, and duplicate papers were omitted. Next, four authors (M.A, N.C, N.A and N.S) screened the titles and abstracts of the remaining studies according to the abovementioned inclusion and exclusion criteria. For the final decision, the full text of the selected studies was assessed. Any uncertainty over the final decision was resolved by an independent expert (E.B).

## 2.6 Data Collection

Four authors (M.A, N.Ch, N.A and N.S) performed the data extraction. They tabulated the data as follows: author and year of publication, procedure (disease severity, prognosis, need for ICU, ventilation requirement, mortality and segmentation), dataset size, age of patients, imaging modality (CT or CXR images), task (classification or segmentation), pre-processing and augmentation of images, model architectures and their performance.

## 2.7 Risk of Bias

The included articles were assessed according to the quality assessment of diagnostic accuracy studies (QUADAS-AI) tool (32), which has been widely used in the AI systematic reviews (33-36). The following domains were used to evaluate the risk of bias (ROB): patient selection, index test, reference standard, and flow and timing. Studies with three or more items with a low risk of bias were considered overall low. Those with only one item at low ROB were evaluated as overall high ROB; others were deemed unclear ROB.

## 2.8 Synthesis of Results and Meta-analysis

The accuracy of the AI models in predicting the need for ventilation, severity assessment, and mortality in COVID-19 patients was determined using Receiver Operator Characteristic (ROC) curves, as evaluated by the AUC value and sensitivity and specificity (true positive (TP), true negative (TN), false positive (FP), false negative (FN) values), if available (37). The meta-analysis included studies that evaluated the sensitivity and specificity of various AI models for predicting the need for ventilation, severity assessment, and mortality in COVID-19 patients. The heterogeneity of included studies was evaluated using I2 and $\chi 2$ statistics and was deemed significant if I2 was more than 50% or the p-value was less than 0.05. To account for the predicted heterogeneity of investigations (38), a random-effects model (DerSimonian-Laird method) was used. Deeks and colleagues, on the other hand, performed a simulation study of tests for publication bias in diagnostic test accuracy (DTA) reviews in 2005 (39). Hence, Deeks' test is suggested and should be preferred for DTA meta-analyses. Furthermore, diagnostic odds ratio (DOR) is defined as the ratio of the chances of testing positive for the target condition to the odds of testing positive without the target condition, (TP/FN):(FP/TN) or LR+/LR- (40). Using STATA version 17 (StataCorp LP, College Station, TX, USA), all plots were generated. In addition, all analyses were conducted using STATA 17.0 software. Accordingly, “midas” and “metandi” were utilized.

# 3 Results

## 3.1 Study Selection

After analyzing the title and abstract of the 1528 studies, the full texts of 193 articles were assessed for eligibility. Ultimately, 36 articles were retained and included for full subsequent analysis (Fig.1).

## 3.2 Study Characteristics

The results of data extraction are presented in Table 3. Among included articles, 24 studies with a total sample size of 358181 examined the severity of the disease (18, 22, 23, 41-61). Amongst them, 15 studies used CT (22, 23, 41, 43, 44, 46, 50-52, 54, 56, 57, 59-61) images and 9 studies used CXR images (24, 42, 45, 47-49, 53, 55, 58). They examined the images of the patients based on the image features, the extent of the infection and lung involvement and then classified the patients into two (45, 47, 48, 53-55, 59, 61, 62), three (22, 41, 56), four (46, 50, 52, 58) or five (51, 57) groups. Three studies differentiated only critical patients admitted to ICU or deaths occurring before or after ICU admission (24, 49, 60). Studies used different models for the classification of the severity, including supported vector machine (SVM) (41, 46, 47, 62), random forest (RF) (47, 56), COV-CAF (51), LungDoc (61), COVID-Net CXR-S (48), ResNet-50 and Inception models (42, 45), Siemens healthiness algorithms (22, 60) and different neural networks (24, 43, 44, 46, 49, 50, 52-54, 58, 59).

Nine studies examined the need for ICU or mechanical ventilation based on CT (8, 61, 63) or CXR (25, 64-68) images with a total sample size of 8239 patients. They reported their classification results as a binary outcome (e.g., whether ventilation or ICU was needed or not). They used different DL- or ML-models including, LungDoc (61), Siemense healthcare (8), DenseNet121 (66), balanced random forest (BRF) (67), RF and convolutional neural network (CNN) (25, 63-65).

Nine studies classified patients according to mortality (whether the patients survive or not). Six studies examined CXR (24, 25, 64, 65, 67-69), and three studies examined CT images (63, 70-72). The total sample size of these studies was 18993 patients. The evaluated architecture were Qure.ai Technologies (19), VGG (69), extreme gradient boosting (XGB) (67, 70), RF and neural networks (25, 63, 71)

## 3.3 Risk of bias and applicability

Two included studies were at high risk of bias (46, 50), seven had unclear risk of bias and others were at low risk of bias (Fig. 2). The index test was the most problematic domain. Also, some studies did not mention the time taken to read CT or CXR records.

## Results of Individual Sources of Studies

Determining the prognosis of COVID-19 disease can generally be classified into three groups, disease severity, mechanical ventilation or need for ICU and mortality.

### 3.3.1 Disease severity

Among articles that examined the disease severity, a range of 0.65 – 0.98 was reported for the area AUC. The DOR of the studies included in this category was between 7.3 and 297.6. The best AUC was reported by Irmak et al. (58), which proposed an automated CNN model for severity classification into four groups mild, moderate, severe and critical. The amount of ground glass, consolidation and lung involvement from 3260 chest X-ray images was evaluated in the study. They reported average accuracy of 0.95, a sensitivity of 0.98 and a specificity of 0.96.

The lowest AUC was related to Balaha et al. (44), which used CNN model with normal augmentation to analyze the image features of the 15535 CT images. It was reported that altering the augmentation approach or even eliminating it can increase the AUC significantly.

Among reported accuracies, a range of 0.72 to 0.98 was obtained. The highest accuracy was achieved by Elsharkawy et al. (55). They developed a model named Markov-Gibbs random field (MGRF) to detect the severity of infection (low severity or high severity) using 200 chest x-ray images. They achieved accuracy, sensitivity and specificity of 0.98, 1.00 and 0.97, respectively, by two-fold cross-validation.

An accuracy of 0.72 was reported in two studies by Shan et al. (23) and Cai et al. (56). Shan et al. (23) used SVM for severity classification (severe or non-severe) based on the quantified radiological features, including the percentage of consolidation (POC), the percentage of infection (POI) and mass of infection (MOI), which were extracted from 549 CT scans. The best prediction accuracy was 0.73 and 0.72 when using MOI and POI, respectively. Also, they concluded that the quantified radiological features are more informative than the pneumonia severity index (PSI), which is a clinical prediction rule. Cai et al. (56) built RF models for severity classification into three groups, moderate, severe and critical, using 99 CT scans. The defined model Ⅰ radiomics as moderate vs. (severe + critical) and model Ⅱ radiomics as severe vs. critical and checked RF performance in each model. The AUC, accuracy, sensitivity and specificity in model Ⅰ were 0.82, 0.75, 0.79 and 0.70, respectively and in model Ⅱ were 0.78, 0.72, 0.79 and 0.66, respectively. Also, they concluded that the hybrid models that combined the radiomics features and clinical data had better performance than those using radiomic features merely.

### 3.3.2 Mechanical Ventilation or Need for ICU

Among studies that reported AUC for the performance of the AI structures, a range of 0.68 to 0.98 was obtained. The DOR was between 4.8 and 76.6. The best AUC was related to Aslam et al. (68), in which CXR images of 1508 patients were analyzed with a combination of DL models and explainable artificial intelligence (EAI). The CXR images were segmented based on features such as opacity and patients were classified accordingly. The authors reported an accuracy of 97% for their model.

A range of 0.52 to 0.97 was reported among studies examining model accuracies. The best accuracy was related to the study by Aslam et al. (68) and the lowest accuracy was written by Aljouie et al. (67). They used four classifiers, including linear SVM, RF, Linear Regression (LR), and XGB on 1508 CXR images to classify patients into mechanical ventilation, non-invasive ventilation and no ventilation groups. They also used some techniques, including Synthetic Minority Over-sampling Technique (SMOTE), Adaptive Synthetic (ADASYN) and random under-sampling (RUS), to improve the performance of the models. The best-achieved performance was an accuracy of 0.52 and an AUC of 0.76 for the BRF using X-ray features. Also, the authors reported that combining X-ray features with clinical and laboratory tests showed better performance.

#### *3.3.3* Mortality

Among articles that examined mortality prediction, a range of 0.74 to 0.99 was reported for AUC. The calculated DOR ranged from 2.16 to 22.6. The highest AUC was reported by Aslam et al. (68) based on the CXR images of 1513 patients. The study reported an accuracy of 98%, the highest among the included papers. The lowest accuracy and AUC were reported by Aljouie et al.(67). The examined four classifiers on 1513CXR images for ventilation requirement and mortality. For mortality prediction, XGB + ADASYN had the best performance (AUC of 0.72 and accuracy of 0.71). The accuracy of the included studies for mortality was 0.71 to 0.83.

## 3.4 Synthesis of results

Figure 3 shows the accuracy, sensitivity and specificity of the included studies. The results of meta-analyses are shown in Table 4.

**Mortality**

Four studies which consisted of seven individual AI models were included in the meta-analysis of mortality prediction using AI in COVID-19 patients. The overall sensitivity and specificity of the included studies were 71% (95% CI: 65%, 77%) and 69% (95%CI: 61%, 76%) respectively (Fig 4A). Moreover, funnel plot (Fig 5A) was symmetric and the asymmetry test p-value of 0.19 indicated that there was no evidence of publication bias. The area under the HSROC curve (AUC) was 0.76 (95%CI: 0.72–0.80) (Fig 6A) indicating moderately accurate optical diagnostic performance of AI in predicting mortality. The DOR value for this outcome was 6 with 95% confidence interval of 3-10.

**Severity Assessment**

In the meta-analysis of assessment of severity using AI in COVID-19 patients, nine investigations including 13 different AI models were considered. Overall, the included studies demonstrated a sensitivity of 88% (95%CI: 77%, 94%) and a specificity of 89% (95%CI: 82%, 94%) (Fig 4B). In addition, the funnel plot (Fig 5B) was symmetric, and the asymmetry test p-value of 0.07 suggested that publication bias was not present. The area under the HSROC curve (AUC) was 0.95 (95%CI: 0.92–0.96) (Fig 6B), showing highly accurate optical diagnostic performance of AI in severity assessment. The DOR value was 59 (95% CI: 18-197).

**Need for Ventilation**

Four studies comprising of six AI models were included in the meta-analysis of AI for predicting ventilation requirements in COVID-19 patients. Overall, the studies that were considered showed a pooled sensitivity of 67% (95%CI: 61%, 73%) and a pooled specificity of 89% (95%CI: 75%, 95%) (Fig 4C). In addition, there was no evidence of publication bias as shown by the symmetric funnel plot (Fig 5C) and a p-value of 0.92 for the asymmetry test. The area under the HSROC curve (AUC) was 0.77 (95%CI: 0.73–0.80) (Fig 6C), demonstrating that AI's optical diagnostic performance in predicting the requirement for ventilation was reasonably accurate. The measured DOR was 16 (95% CI: 7-36)

# 4 Discussion

Machine learning and deep learning methods facilitate the extraction and identification of body tissues characteristics from images and thus speed-up patient triage and allow timely treatment plans for patients. Therefore, in the current study, we reviewed the studies that analyzed the performance of AI models for predicting disease severity, ventilation requirement, need for ICU and mortality using standard of care CT or CXR images.

CXR and CT imaging modalities were used in the included studies. Chest radiography is a quick and easy test and is usually requested due to low cost and fast data acquisition compared to CT (73). However, it was reported that CXR has restrictions for the accurate detection of COVID-19 infection compared to CT. On the other hand, CT images are better options for disease severity analysis and patient monitoring, and they have shown higher sensitivity compared to CXR (41). Another possible source of bias from CXR is that AI methods may evaluate images taken from different views leading to inaccurate outcome (21). For instance, instead of posteroanterior view, in severe cases an anteroposterior projection is used. Having mentioned the above points, based on our results, machine learning models that were applied to CXR showed accuracy of 95-98% had comparable accuracy with CT (i.e., 72-97 % based on 10 studies) to evaluate the severity of the disease. However, there was no study aimed at comparing the results of these two data acquisition methods.

The biggest limitation behind using CT and CXR images for diagnosis and evaluating the prognosis of the disease is the lack of COVID-related experience among radiologists concerning the COVID-19 infection pathways and spread. Besides, there is always the possibility of error when human vision is used to analyze the images. In the early stage of the pandemic, the progression patterns of the disease were not completely recognized and showed different behaviors in each region. Besides, considering the variations in the health and triage system in different regions, data regarding the virus behavior in one region cannot be generalized to all countries. Initially, due to decrease the errors, it was recommended to design scoring systems to evaluate images objectively. This has resulted in more accurate decision-making and increased efficacy. However, manual segment scoring is still time-consuming and may not be optimal for daily clinical practice. Thus, AI-based methods have the potential to decrease workload and improve patient safety (74).

In the included studies, the severity of the COVID-19 infection severity was assessed using different approaches. One of the most common methods was whole lungs/lesions segmentation and evaluation based on the extent of the affected tissue. Most included studies used UNet models to segment the lungs and lesion areas. They similarly obtained a dice similarity coefficient (DSC) of about 0.98 for lesion segmentation with the range of 0.77-0.94. Li, Z et al. (50) used a feature pyramid network (FPN) to achieve the best DSC. They reported that although FPN did not improve the results compared with the UNet model in lung segmentation, it showed better results in lesion segmentation. The lowest DSC was reported by Cai et al. (56) using the UNet model. Compared to the study performed only lung assessment (57), all studies that performed both lung and lesion segmentation had higher accuracy in severity assessment except one (56).

Studies used different categorizing methods to classify the severity among the patients (75). Having more classes will increase the precision of the patient categorization and will improve the treatment response (42). However, this can complicate the data processing, which can decrease the model performance. Most of the studies that analyzed the severity level categorized patients in two groups, and the best performance belonged to one of these models.

It must be noted that factors such as age, sex and body mass can impact the response of human body to the infection. Indeed, patients with different abovementioned features may have different prognosis even with the same initial infection stage. Thus, neglecting these variables in evaluating the function of the AI models in some of the included studies can be considered as a significant drawback. This issue is even more important in studies that evaluate more advanced outcomes, such as the need for ventilation or intensive care and mortality (43).

The AI model learn from the available training data. Thus, a non-curated data could include some inaccuracies; hence lower the performance of the AI model. Therefore, a more reliable outcome can be expected in case of analyzing models in which the ground truth is accurate and reliable since the machine is trained based on the imported ground truth. The gold standard may be less accurate in the case of analyzing severity with AI models since it depends on the practitioner assessment, which may differ from site to site and expert from expert. On the other hand, as mortality and the need for ventilation are variables that have a binary condition (i.e., will happen or not), the gold standard of the models developed for these two variables can be considered ground truth which is a critical advantage for this model.

In clinical routine practice, AI methods can accelerate the triage, aid decision makers in stressful situations and enable practitioners to help people in a broader area. It is recommended that based on their scoring system, the necessity of ventilation, intensive care, and the possibility of mortality in each of the mentioned situations can be discussed with patient, which can significantly help them in decision making (43). Besides, this scoring system can estimate the length of stay and the duration of high-level of care.

## Limitations

Public databases of CT and CXR images of patients with COVID-19 provide a valuable source for AI research. Although these studies have been performed at different institutions across the globe, almost all AI systems are not open and unavailable for the research community. Besides, in order to evaluate generalizability of the models, individual participant data (IDP) from different regions can be used which can significantly increase the applicability and robustness of the models in daily routine care (76, 77). Accordingly, the world health organization has designed a platform for sharing anonymized COVID-19 clinical data (3).

In case of combining public data sets to train or test the model, it has to be considered that most of them have no restrictions on the imported data, hence the possibility of using duplicated images or even those that are not correctly diagnosed with COVID-19. Besides, since all of the images are not in the DICOM format, a decreased image quality can be expected (21). This can cause a serious problem for machine learning models since the amount of decrease is not the same among the images. Neglecting the demographics of patients and adding pediatric images in public data sets have increased the bias of using them in analyses (21).

Moreover, included studies did not provide complete data concerning the function of their models such as sensitivity, specificity and accuracy. This prevented other research from reproducing and comparing the achieved performance. Image modalities that used in different studies had different features and setups. Despite, the extensive efforts in developing ML models using different feature extraction, selection and classification algorithms and DL models using different architecture and topologies; the comparison of their performances and applicability is, at least, challenging at this stage.

Imaging scans were acquired at various institutions using different scanners and data acquisition and image reconstruction protocols. Accordingly, the obtained images should be pre-processed to  ensure consistency of the input (78). Imaging systems and scanning protocols for acquiring images use different acquisition parameters, and so are CT image reconstruction methods. These factors can significantly impact the robustness and reliability of AI applications and lead to misdiagnosis.

Included studies lack an independent external validation. Thus, although the majority of the included studies were at low risk of bias, it should be noted that we cannot recommend any model to be used in daily practice, specifically considering that recent publications about COVID-19 prediction models are entering the literature quickly.

Furthermore, several studies reviewed here did not mention the imaging study duration despite being an essential factor in determining the prognosis and timely determination of the prognosis leads to appropriate treatment. Although most images are acquired during admission, it has not been evaluated whether the models will have the same predictive values about the need for intensive care or mortality if images are taken at other time points.

In order to use prediction models for decision-making, it is essential for the studies to assess the performance of a diagnostic tool to specify the target population, enabling users to know which category of patients can be evaluated using a given model (3). However, this data was not comprehensibly provided in the included studies, which made users doubt whether to use the model for their intended population. Considering the variabilities in the target population can justify the discrepancies in the results reported by different studies, the difference in the relative frequency between the population necessitates some alternations in the prediction model are required in other settings (3).

With regards to the choice of predictors in the prediction model, it is recommended to consider the expert opinion and published literature rather than choosing only the data-driven ones. For prediction models, the following variables are recommended: age, sex, C reactive protein, lactic dehydrogenase, lymphocyte count, CT-scoring, albumin (or albumin/globin), direct bilirubin and red blood cell distribution width.

Despite all recent progress, the proposed methods commonly face challenges for implmenetation in routine practice due to the following reasons: (1) the bias due to small datasets; (2) the variations observed in large internationally sourced datasets; (3) the poor integration of multistream data, particularly imaging data; (4) the difficulty of the task of prognosis; and (5) the necessity for clinicians and data analysts to work together to ensure the developed algorithms are clinically relevant and applicable into routine clinical care. Overall, there is a significant need for creation of trustworthy ecosystems towards routine deployment of AI techniques (79)

## Conclusion

Machine learning and deep learning models can help clinicians predict the severity of disease, ventilation requirement or need for ICU, and mortality, and to subsequently manage COVID-19 patients more effectively. Based on evidence from included studies, the models using imaging data extracted from CT or CXR reported adequate levels of performance. However, the proposed methods commonly face challenges for deployment in routine practice due to issues concerning data curation, harmonization of imaging protocols, reproducibility, external validation, explainability, robustness and applicability, and overall lack of following best practices in AI development and validation. Furthermore, it is essential to provide statistical analysis of model performances, including sensitivity, specificity and accuracy, enabling researchers to compare models more objectively.

## Acknowledgements and Compliance with Ethical Standards

**Competing of interest**: The authors have no relevant financial or non-financial interests to disclose.

**Funding**: No funding was received to assist with the preparation of this manuscript.

**Research involving Human Participants and/or Animals**: Not applicable

**Informed consent**: Not applicable

**Conflict of Interests:**

**SaeedReza Motamedian:** The authors have no relevant financial or non-financial interests to disclose.

**Sadra Mohaghegh:** The authors have no relevant financial or non-financial interests to disclose.

**Elham Babadi Oregani:** The authors have no relevant financial or non-financial interests to disclose.

**Mahrsa Amjadi:** The authors have no relevant financial or non-financial interests to disclose.

**Parnian Shobeiri:** The authors have no relevant financial or non-financial interests to disclose.

**Negin Cheraghi:** The authors have no relevant financial or non-financial interests to disclose.

**Niusha Solouki:** The authors have no relevant financial or non-financial interests to disclose.

**Nikoo Ahmadi:** The authors have no relevant financial or non-financial interests to disclose.

**Hossein Mohammad-Rahimi:** The authors have no relevant financial or non-financial interests to disclose.

**Yassine Bouchareb:** The authors have no relevant financial or non-financial interests to disclose.

**Arman Rahmim:** The authors have no relevant financial or non-financial interests to disclose.

# Figure Legends

**Figure 1.** PRISMA flow diagram (literature search strategy and study selection).

**Figure 2**. Risk of bias of the included studies.

**Figure 3.** Distribution of the specificity, sensitivity and accuracy of the included studies categorized based on the study aim.

**Figure 4**. Forest plot of sensitivity and specificity of AI in predicting mortality (A), severity assessment (B) and predicting the need for ventilation (C)

**Figure 5.** Deeks' funnel plot to evaluate publication bias of studies in predicting mortality (A), severity assessment (B) and predicting the need for ventilation (C) . The vertical axis displays the inverse of the square root of the effective sample size (1/root(ESS)). The horizontal axis displays the diagnostic odds ratio (DOR). All p values indicated a symmetrical funnel plot.

**Figure 6.** Hierarchical summary receiver-operating characteristic (HSROC) curve for the diagnostic performance of AI in predicting mortality (A), severity assessment (B) and predicting the need for ventilation (C). The size of the gray circles indicates the number of samples in the individual studies. The summary sensitivity and specificity are shown with a dark red square and the 95% confidence region is plotted in short lines.

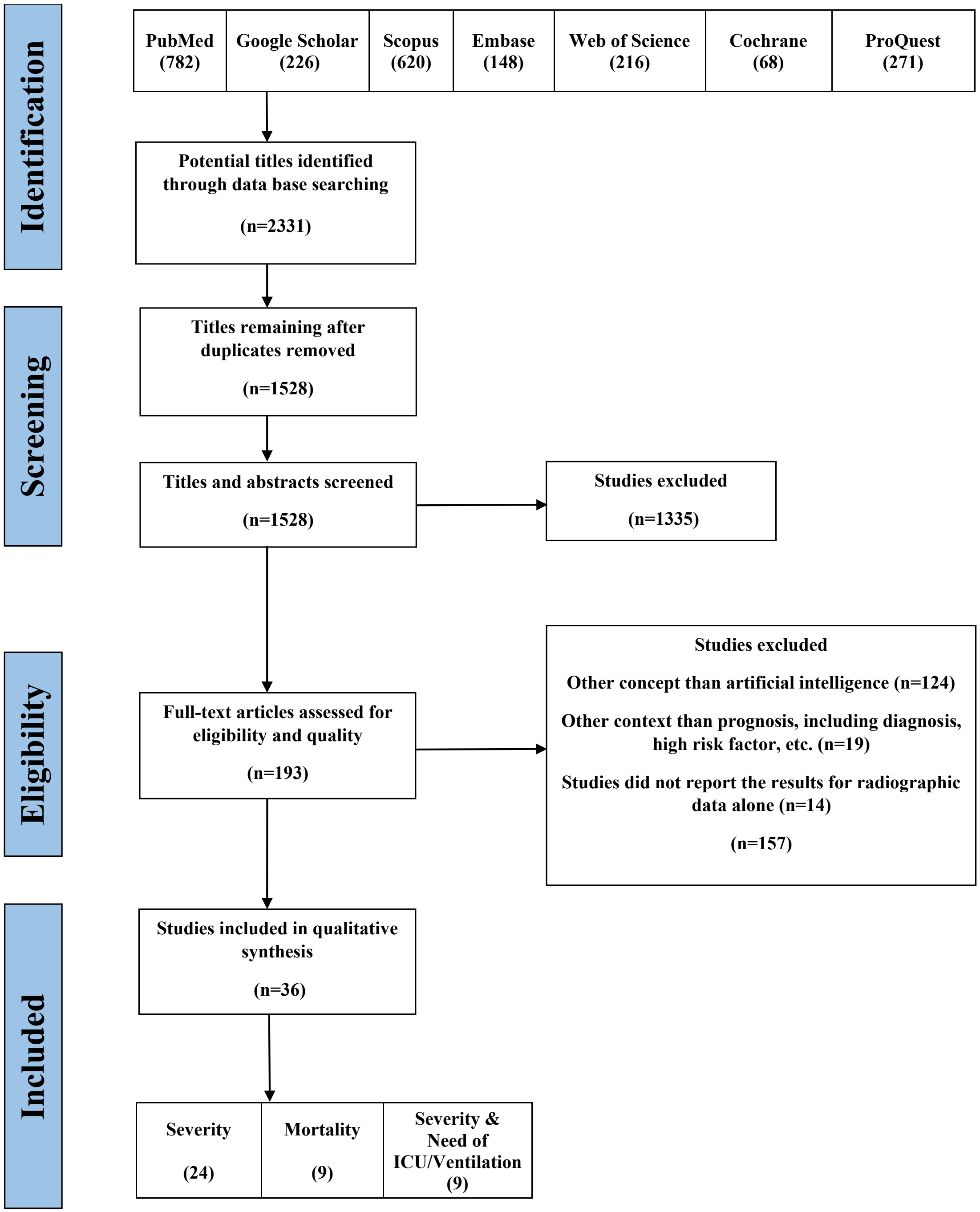

**Figure 2.** PRISMA flow diagram (literature search strategy and study selection).

| | Patient Selection | Index Test | Reference Standard | Flow and Timing | Overall |
|---|---|---|---|---|---|
| **Shalbaf 2022** | Low | High | High | Unclear | High |
| **Li, Z 2021** | High | High | Low | Unclear | High |
| **Ahmad 2022** | Low | High | Low | Low | Low |
| **Aljouie 2021** | Low | High | Low | Low | Low |
| **Shan 2021** | Low | High | Low | Low | Low |
| **Jiao 2021** | Low | High | Low | Low | Low |
| **Lassau 2021** | Low | High | Low | Low | Low |
| **Li, M 2020** | Low | High | Low | Low | Low |
| **Gieraerts 2020** | Low | High | Low | Low | Low |
| **Gouda 2020** | High | Low | Low | Low | Low |
| **Cai 2020** | High | Low | Low | Low | Low |
| **Li, Y 2020** | High | Low | Low | Low | Low |
| **Ahmed T 2022** | Low | Low | Low | Low | Low |
| **Aslam 2022** | Low | Low | Low | Low | Low |
| **Bermejo 2022** | Low | Low | Low | Low | Low |
| **Chamberlin 2022** | Low | Low | Low | Low | Low |
| **Jordan 2022** | Low | Low | Low | Low | Low |
| **Munera 2022** | Low | Low | Low | Low | Low |
| **Spagnoli 2022** | Low | Low | Low | Low | Low |
| **Bae 2021** | Low | Low | Low | Low | Low |
| **Ho 2021** | Low | Low | Low | Low | Low |
| **Purkayastha 2020** | Low | Low | Low | Low | Low |
| **Balaha 2022** | Low | Low | Low | Unclear | Low |
| **Ortiz 2022** | Low | Low | Low | Unclear | Low |
| **Shiri 2022** | Low | Low | Low | Unclear | Low |
| **Kulkarni 2021** | Low | Low | Low | Unclear | Low |
| **Shiri 2021** | Low | Low | Low | Unclear | Low |
| **Qiblawey 2021** | Low | Low | Low | Unclear | Low |
| **Irmak 2021** | Low | Low | Low | Unclear | Low |
| **Kohli 2021** | High | Low | Low | High | Unclear |
| **Mushtaq 2020** | High | High | Low | Low | Unclear |
| **Abbasi 2022** | Low | High | Low | Unclear | Unclear |
| **Aboutalebi 2022** | Low | High | Low | Unclear | Unclear |
| **Dinh 2022** | Low | High | Low | Unclear | Unclear |
| **Elsharkawy 2021** | Low | High | Low | Unclear | Unclear |
| **Ibrahim 2021** | High | Low | Low | Unclear | Unclear |

**Figure 2**. Risk of bias of the included studies.

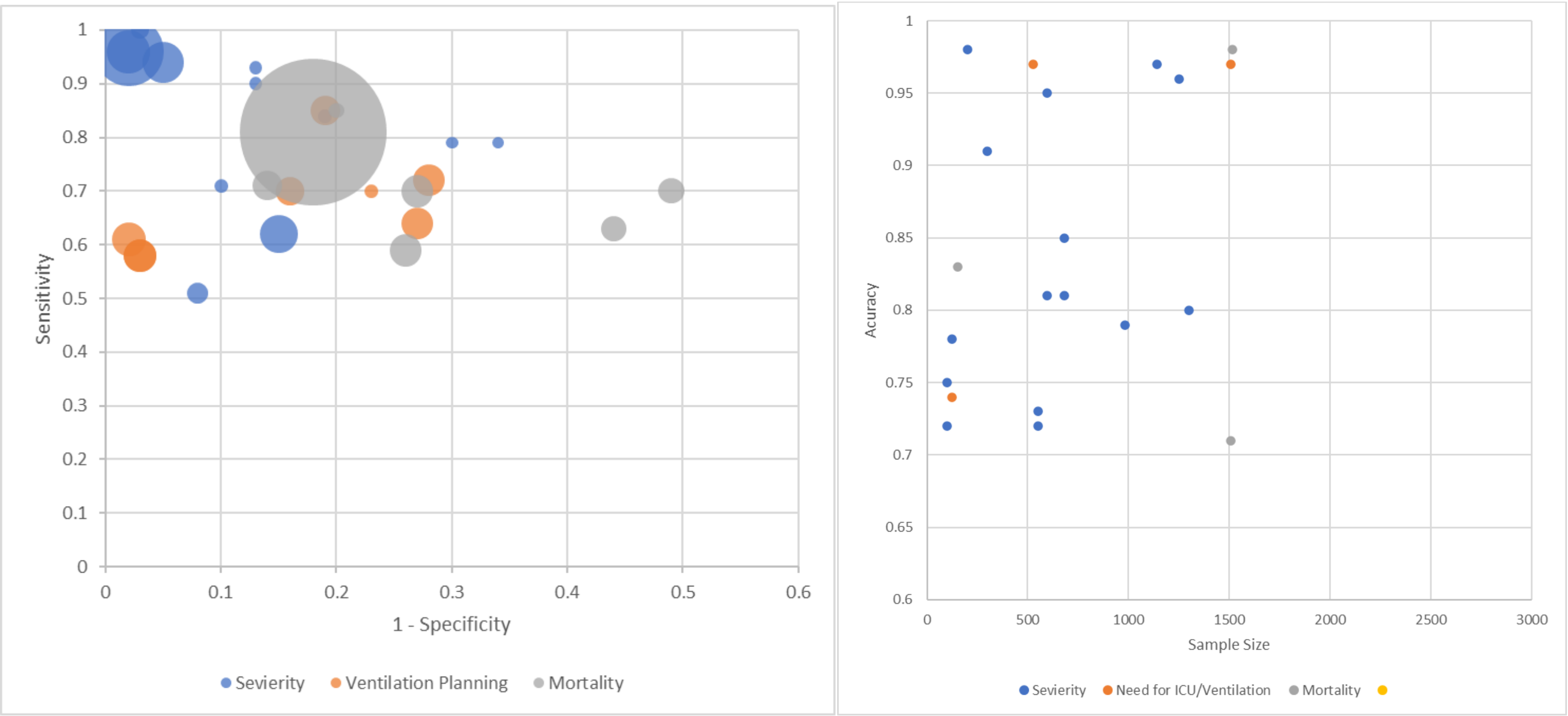

**Figure 3.** Distribution of the specificity, sensitivity and accuracy of the included studies categorized based on the study aim.

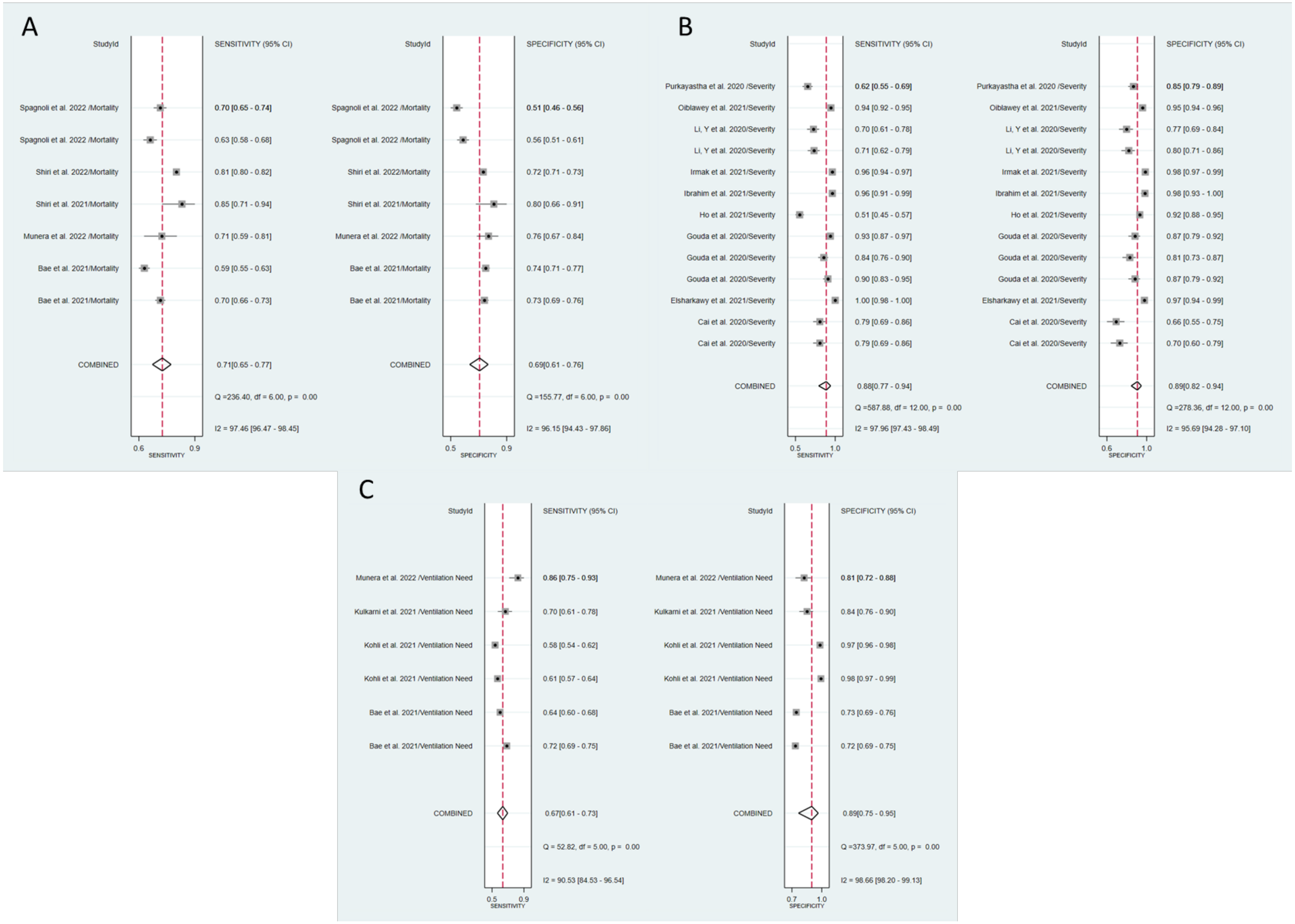

**Figure 4**. Forest plot of sensitivity and specificity of AI in predicting mortality (A), severity assessment (B) and predicting the need for ventilation (C)

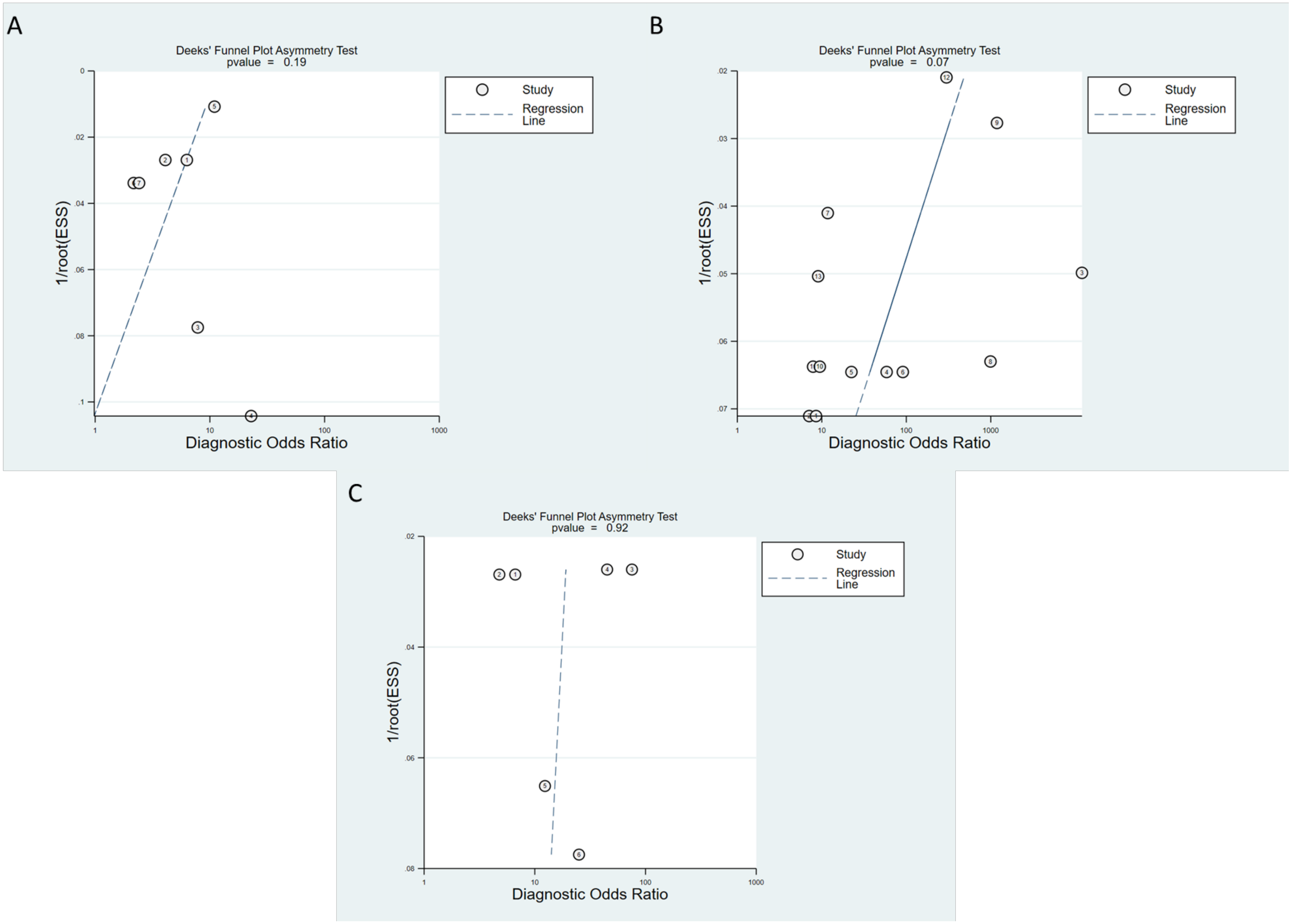

**Figure 5.** Deeks' funnel plot to evaluate publication bias of studies in predicting mortality (A), severity assessment (B) and predicting the need for ventilation (C) . The vertical axis displays the inverse of the square root of the effective sample size (1/root(ESS)). The horizontal axis displays the diagnostic odds ratio (DOR). All p values indicated a symmetrical funnel plot.

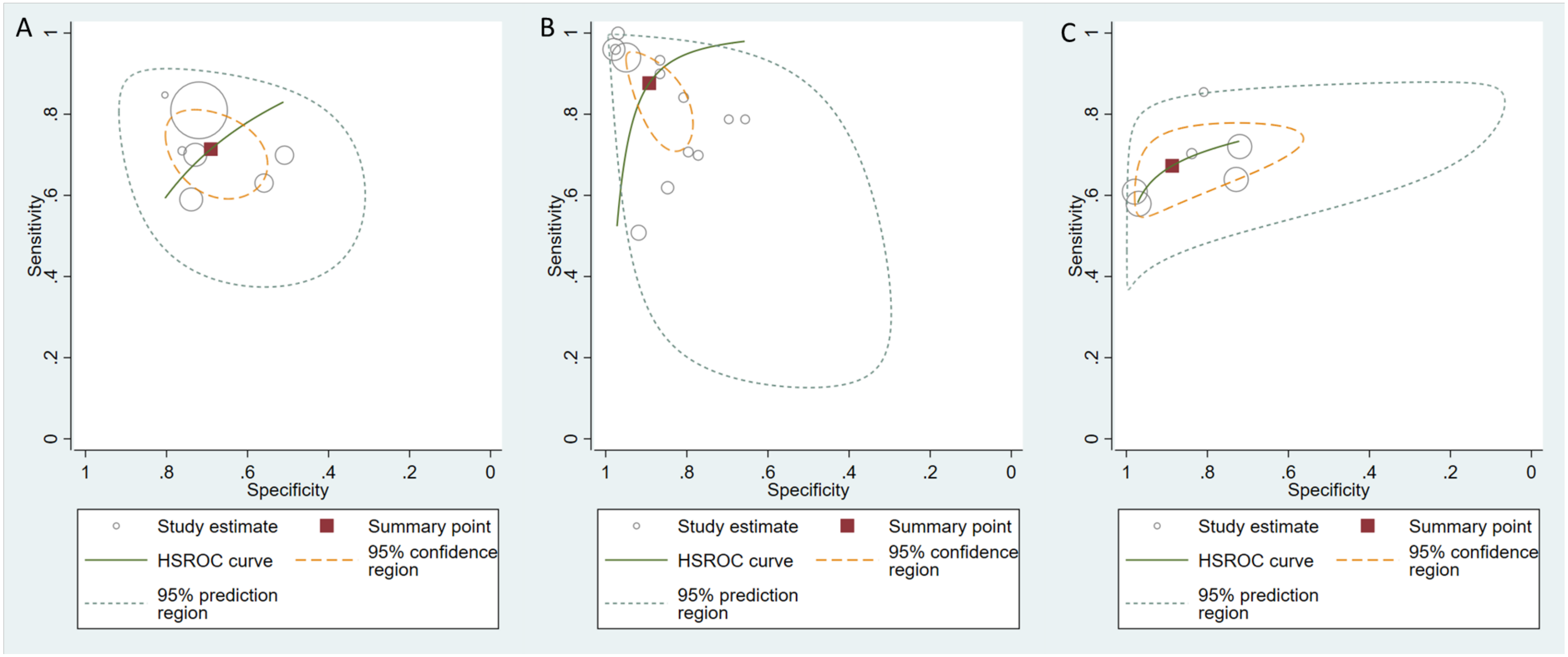

**Figure 6.** Hierarchical summary receiver-operating characteristic (HSROC) curve for the diagnostic performance of AI in predicting mortality (A), severity assessment (B) and predicting the need for ventilation (C). The size of the gray circles indicates the number of samples in the individual studies. The summary sensitivity and specificity are shown with a dark red square and the 95% confidence region is plotted in short lines.

**Table 1**. Inclusion and exclusion criteria used for study selection.

| **Table 1. Inclusion and Exclusion Criteria** | | |
|---|---|---|
| | **Inclusion Criteria** | **Exclusion Criteria** |
| **Population= Covid-19** | Studies analyzing patients suffering from COVID-19 infection. | None |
| **Intervention= AI** | Literature used artificial intelligence, deep learning or machine learning techniques based on radiographic images, including CXR and CT images. | If the results were not reported merely based on radiographic images and combined with clinical and laboratory information. |
| **Comparison= Gold standard (actual condition of the patients)** | None | Studies that did not specify the ground truth |
| **Outcome: Prognosis** | Studies were performed to determine the severity, prognosis, recurrence, mortality and survival rate of the COVID-19disease. Also, studies that assessed the treatment outcomes were included. | None |

**Table 2**. Queries used for the electronic search in different data bases.

| Table 2. Search Queries | | |
|---|---|---|
| **Motor Engine** | **Search Query** | **Result** |
| **PubMed** | ("artificial intelligence"[MeSH] OR "AI" OR "machine learning"[MeSH] OR "ML" OR "deep learning"[MeSH] OR "DL" OR "big data"[MeSH] OR "computer aided" OR ""diagnosis, computer assisted"[MeSH Terms]" OR "neural network") AND ("COVID-19"[MeSH] OR "SARS-CoV-2"[MeSH] OR "coronavirus"[MeSH] OR "covid-19" OR "sars-cov-2" OR "coronavirus") AND ("prognosis"[MeSH] OR "mortality"[MeSH] OR "prognostic" OR "prediction" OR "severity" OR predict OR "Treatment Outcome"[Mesh] OR mortality OR survival OR recurrence) | 782 |
| **Google Scholar** | ("artificial intelligence" OR "AI" OR "machine learning" OR "deep learning" OR "big data" OR "computer aided") AND ("COVID-19" OR "SARS-CoV-2" OR "coronavirus") AND ("prognosis" OR "prognostic" OR "severity") | 226 |
| **Scopus** | ("artificial intelligence" OR "AI" OR "machine learning" OR "ML" OR "deep learning" OR "DL" OR "big data" OR "computer aided" OR "diagnosis, computer assisted" OR "neural network") AND ("COVID-19" OR "SARS-CoV-2" OR "coronavirus") AND ("prognosis" OR "prognostic" OR "prediction" OR "severity" OR “predict” OR "Treatment Outcome" OR “mortality” OR “survival” OR “recurrence”) | 620 |
| **Embase** | ("artificial intelligence" OR "AI" OR "machine learning" OR "ML" OR "deep learning" OR "DL" OR "big data" OR "computer aided" OR "diagnosis, computer assisted" OR "neural network") AND ("COVID-19" OR "SARS-CoV-2" OR "coronavirus") AND ("prognosis" OR "prognostic" OR "prediction" OR "severity" OR “predict” OR "treatment Outcome" OR “mortality” OR “survival” OR “recurrence”) | 148 |
| **Web of Science** | (‘artificial intelligence’ OR ‘ai’ OR ‘machine learning’ OR ‘deep learning’) AND (‘covid-19’ OR ‘sars-cov-2’ OR ‘coronavirus’) AND (‘prognosis’ OR ‘severity’ OR ‘mortality’) | 216 |
| **Cochrane** | ("artificial intelligence" OR “AI” OR "machine learning" OR "deep learning" OR “DL” OR "big data" OR "computer aided" OR "neural network") AND ("COVID-19" OR "SARS-CoV-2" OR "coronavirus") AND ("prognosis" OR "prognostic" OR "predict" OR "Treatment Outcome" OR “mortality” OR “survival” OR “recurrence”) | 68 |
| **ProQuest** | SU.X("deep learning") AND "covid 19" AND "prognosis" | 271 |

**Table 3**. Results of data extraction.

| Author, Year | **Desired outcome** | **Dataset Size (train/ test), Sex** | **Age** | **Imaging Modality** | **Task** | **Pre-Processing (P) Augmentation (A)** | **Model Architecture** | **Parameters/Features** | **Reported Prognostic performance** | **Performance** |
|---|---|---|---|---|---|---|---|---|---|---|
| Ortiz 2022 (41) | Severity (mild, moderate and severe) | 596 Test: 596 | *NA* | CT | Classification | *NA* | SVM | Infection histogram | Accuracy | 0.81 |
| | | | | | | | SVM with Latent Dirichlet Alocation (LDA) | | | 0.95 |
| Dinh 2022 (42) | Severity | 2629 (2479/150) | NA | CXR | Classification | P:NA A: height_shift_range, rotation_range, horizontal_flip, brightness_range, width_shift_range, and rescale | DenseNet121 | Image features | Micro-average | 0.81 |
| | | | | | | | ResNet50 | | | 0.76 |
| | | | | | | | InceptionNet | | | 0.75 |
| | | | | | | | Swin Transformer | | | 0.80 |
| | | | | | | | Hybrid EfficientNet-DOLG | | | 0.82 |
| Chamberlin 2022 (43) | Severity | 93 M= 57 F= 36 | 59 | CT | segmentation | P: + A: NA | dCNN | Lesion region | Opacity score | 0.89 |
| Balaha 2022 (44) | Severity | 15535 (5.159/10.376) | NA | CT | classification | *No augmentation | CNN | Image features | AUC | 0.99 |
| | | | | | | Normal augmentation | | | | 0.65 |
| | | | | | | CC-GAN | | | | 0.99 |
| | | | | | | CycleGAN | | | | 0.99 |
| Ahmad 2022 (45) | Severity (Improved, Deterioration) | 877 (582/295) | NA | CXR | classification | NA | RresNet-50 | Severity score | AUC | 0.85 |
| | | | | | | | InceptionV3 | | | 0.88 |
| | | | | | | | InceptionResNet V2 | | | 0.89 |
| | | | | | | | ChexNet | | | 0.92 |
| | | | | | | | EfficientNet | | | 0.87 |
| Shalbaf 2022 (46) | Severity (normal, mild, moderate and severe) | 683 Train: 547 Validation: 68 Test: 68 F: 275 M:408 | *NA* | CT | Classification | P: extracting the central region of the images, and converting to binary images A: *NA* | An ensemble of five pre-trained CNN models with SVM | Image features | Accuracy | 0.81 |
| | | | | | | | An ensemble of five pre-trained CNN models with fine-tuning and softmax | | | 0.85 |
| Abbasi 2022 (47) | Severity prediction (More or less severe) | 278 (114/164) | *NA* | CXR | Classification | P: image resizing (313 × 313 pixels), de-noising, and contrast stretching A:NA | SVM | Severity Score | AUC | 0.96 |
| | | | | | | | | | F1 | 0.90 |
| | | | | | | | RF* | | AUC | 0.90 |
| | | | | | | | | | F1 | 0.84 |
| | | | | | | | XGBoost | | AUC | 0.96 |
| | | | | | | | | | F1 | 0.90 |
| Aboutalebi 2021 (48) | Severity (two levels) | 258 (208/150) M:161 F:97 | 59.11 | CXR | segmentation | P: Cropped, resampled to 480*480. | COVID-Net CXR-S | Opacity | Sensitivity | 0.92 |
| | | | | | | | | | Sensitivity | 0.92 |
| | | | | | | | | | Sensitivity | 0.92 |
| | | | | | | A: Translation, rotation, horizontal flip, zoom, and intensity shift | | | | |
| Gouda 2020 (22) | Severity (mild, severe, critical) | 120<br>Test: 120<br>F: 22<br>M: 98 | 52.63 | CT | Classification | *NA* | Siemens Healthineers | Total severity score | AUC | 0.94 |
| | | | | | | | | | Sensitivity | 0.90 |
| | | | | | | | | | Specificity | 0.87 |
| | | | | | | | | Total opacity score | AUC | 0.92 |
| | | | | | | | | | Sensitivity | 0.84 |
| | | | | | | | | | Specificity | 0.81 |
| | | | | | | | | Total score for crazy paving and consolidation | AUC | 0.97 |
| | | | | | | | | | Sensitivity | 0.93 |
| | | | | | | | | | Specificity | 0.87 |
| Li, Z 2021 (50) | Lung and lesion segmentation | Lung: 5750CT<br>Lesion:1117CT<br>(train/test= 4:1) | *NA* | CT | Segmentation | P: Lung segmentation, Lesion segmentation, 3D visualization<br>A: Left–right flip, Top–bottom flip, Top–bottom and Left–right flip, ± 15-degree rotation, ± 30-degree rotation with 12.5% probability | U-net++ | Lung regions | DSC | 0.97 |
| | | | | | | | | Lesion regions | | 0.84 |
| | Severity (early, progressive, peak, absorption) | 1301<br>Train: 1035<br>Test: 266 | | | Classification | | Dual-Siamese channels and Clinical metadata | Image features | Accuracy | 0.80 |
| Shan 2021 (23) | Lesion segmentation | 549<br>Train: 249<br>Test: 300<br>F: *NA*<br>M: *NA* | >18 | CT | Segmentation | *NA* | VB-Net neural network | Lesion regions | DSC | 0.91 |
| | Severity (severe or non-severe) | | | | Classification | | SVM, C-SVM | MOI | Accuracy | 0.73 |
| | | | | | | | | POI | | 0.72 |
| Ibrahim 2021 (51) | Severity | 1252<br>Train: 1126<br>Test: 126<br>F: *NA*<br>M: *NA* | *NA* | CT | Classification | P: Converting the 3D-CT volumes to 2D-slices<br>A: *NA* | A novel computer aided framework (COV-CAF) | Amount of ground glass, consolidation and lung involvement | AUC | 0.97 |
| | | | | | | | | | Accuracy | 0.96 |
| | | | | | | | | | Sensitivity | 0.96 |
| | | | | | | | | | specificity | 0.98 |
| Qiblawey 2021 (52) | Lung segmentation | 1139<br>F: *NA*<br>M: *NA* | *NA* | CT | Segmentation | P: Normalizing the image intensity values and mapping to pixel values, changing the intensity interval to create consistent image and resizing | DenseNet 121 UNet | Lung region | DSC | 0.97 |
| | Lesion segmentation | | | | | | DenseNet 201 FPN | Lesion region | | 0.94 |
| | Severity | | | | Classification | | ED-CNN | Infection percentage | Accuracy | 0.97 |
| | (mild, moderate, severe, critical) | | | | | the images to 256×256<br>A: applying rotations of 90, −90, 180 degrees for CT images and ground truth masks | | | Sensitivity | 0.94 |
| | | | | | | | | | Specificity | 0.95 |
| Irmak 2021 (58) | Severity (mild, moderate, severe & critical) | 3260 images<br>Train: 1956<br>Vlidation: 652<br>Test: 652 | *NA* | CXR | Classification | *NA* | CNN | Opacity degree and lung involvement | AUC | 0.98 |
| | | | | | | | | | Accuracy | 0.95 |
| | | | | | | | | | Sensitivity | 0.96 |
| | | | | | | | | | Specificity | 0.98 |
| Jiao 2021 (53) | Severity (critical or non-critical) | 1834<br>Train: 1285<br>Validation: 183<br>Test: 366<br>F:1177<br>M:1132 | 56 | CXR | Classificatin | P: Normalizing CXRs to the range 0–1<br>A: *NA* | EfficientNet-B0 | Image features | AUC | 0.75 |
| Lassau 2021 (54) | Severity (low- moderate- or high-risk) | 1626<br>Train: 646<br>Test: 980 | 62.6 | CT | Classification | P: Resizing the CT scans to a fixed pixel spacing of (0.7 mm, 0.7 mm, 10 mm), Applying a specific windowing on the HU intensities<br>A: *NA* | A neural network containing two submodel: Resnet50 EfficientNet B0 | Infection percentage | AUC | 0.76 |
| Elsharkawy 2021 (55) | Severity (low severity or high severity) | 200<br>Test: 200<br>F:*NA*<br>M:*NA* | *NA* | CXR | Classification | P: Segmentation of lung region, Enhancement of contrast, Extracting candidate of abnormal tissues<br>A: Scale, Rotation, Translation | MGRF and a neural network based fusion system | Gibbs energy CDF | Accuracy | 0.98 |
| | | | | | | | | | Sensitivity | 1.00 |
| | | | | | | | | | Specificity | 0.97 |
| Cai 2020 (56) | Lung and lesion segmentation | 99<br>Test: 99<br>F: 41<br>M:58 | 54.5 | CT | Segmentation | P: Lung CT window level setting, Batch normalization<br>A: Adding noise, Random rotation, Random shift, Random shear, Random zoom | U-Net | Lung region | DSC | 0.98 |
| | | | | | | | | Lesion region | | 0.77 |
| | Severity (model Ⅰ: moderate vs severe+critical, model Ⅱ: severe vs critical) | | | | Classification | | RF | Lung involvement moderate vs severe+critical | AUC | 0.82 |
| | | | | | | | | | Accuracy | 0.75 |
| | | | | | | | | | Sensitivity | 0.79 |
| | | | | | | | | | Specificity | 0.70 |
| | | | | | | | | Lung involvement severe vs critical | AUC | 0.78 |
| | | | | | | | | | Accuracy | 0.72 |
| | | | | | | | | | Sensitivity | 0.79 |
| | | | | | | | | | Specificity | 0.66 |
| Purkayastha 2020 (57) | Lung segmentation | 981<br>Train: 784<br>Tesy: 197<br>F: 475<br>M: 506 | 48.9 | CT | Segmentation | *NA* | A deep convolutional neural network | Lung region | DSC | *NA* |
| | Severity | | | | Classification | | TSCR + KNN | Image features | AUC | 0.74 |
| | | | | | | | | | Accuracy | 0.79 |
| | | | | | | | | | Sensitivity | 0.62 |
| | | | | | | | | | Specificity | 0.85 |
| Ho 2021 (59) | Severity (high risk or low risk) | 297<br>Test:297<br>F:169<br>M:128 | 60 | CT | Classification | P: Lung segmentation, Removing background, Lesion classification<br>A: *NA* | CNN | Image features | AUC | 0.81 |
| | | | | | | | | | Accuracy | 0.91 |
| | | | | | | | | | Sensitivity | 0.51 |
| | | | | | | | | | Specificity | 0.92 |
| | | | | | | | | | TN, FN, TP, FP | 50, 4, 3, 0 |
| Gieraerts 2020 (60) | Severity (critical patient) | 250<br>Test: 250<br>F:133<br>M:117 | 66.6 | CT | Segmentation Classification | *NA* | Siemens Healthineers | Lung involvement | AUC | 0.87 |
| | | | | | | | | CT score | | 0.88 |
| Li, M 2020 (49) | Severity (critical patient) | 468<br>Train:314<br>Test: 154<br>F: 276<br>M: 192 | 57 | CXR | Classification | *NA* | Siamese neural network–based algorithm | Severity score | AUC | 0.80 |
| Li, Y 2020 (61) | Severity (non-severe and in progress-to-severe) | 123<br>Test: 123<br>F:61<br>M:62 | 64.43 | CT | Classification | *NA* | LungDoc | Consolidation volume | AUC | 0.79 |
| | | | | | | | | | Accuracy | 0.78 |
| | | | | | | | | | Sensitivity | 0.71 |
| | | | | | | | | | Specificity | 0.80 |
| | Need of ICU | | | | | | | | AUC | 0.75 |
| | | | | | | | | | Accuracy | 0.74 |
| | | | | | | | | | Sensitivity | 0.70 |
| | | | | | | | | | Specificity | 0.77 |
| Mushtaq 2020 (24) | Severity (critical patient) | 697<br>Test: 697<br>F: 232<br>M: 465 | 62 | CXR | Classification | *NA* | qXR,v2.1 c2, Qure.ai Technologies | Lung involvement | AUC | 0.77 |
| | Mortality | | | | | | | | | 0.66 |
| Kohli | Ventilation requirement | 740 | 59 | CT | Segmentation | *NA* | | OS1 | AUC | 0.92 |
| 2021 (8) | | Test: 740<br>F : 257<br>M: 482 | | | Classification | | Siemens Healthcare version 2.5.2 | | Sensitivity | 0.61 |
| | | | | | | | | | Specificity | 0.98 |
| | | | | | | | | OS2 | AUC | 0.91 |
| | | | | | | | | | Sensitivity | 0.58 |
| | | | | | | | | | Specificity | 0.97 |
| | | | | | | | | OP | AUC | 0.91 |
| | | | | | | | | | Sensitivity | 0.58 |
| | | | | | | | | | Specificity | 0.97 |
| Kulkarni<br>2021 (66) | Ventilation requirement | 528<br>Train: 410<br>Test: 118<br>F:170<br>M:358 | 54.4 | CXR | Classification | P: Resizing to 224×224 pixels, Centre cropped<br>A: Random combination of right or left rotation (maximum 30°), Random cropping, Random lighting | DenseNet121 | Image features | AUC | 0.97 |
| | | | | | | | | | Accuracy | 0.97 |
| | | | | | | | | | Sesitivity | 0.70 |
| | | | | | | | | | Specificity | 0.84 |
| | | | | | | | | | TN, FN, TP, FP | 92, 13, 30,18 |
| Munera<br>2022 (64) | Need of ICU | 582<br>Train: 408<br>Validation: 105<br>Test: 69<br>F: *NA*<br>M: *NA* | *NA* | CXR | Classification | P: take all images to the same dynamic range and remove elements that were not part of the image<br>A: *NA* | CNN | Image features | AUC | 0.88 |
| | | | | | | | | | Sensitivity | 0.85 |
| | | | | | | | | | Specificity | 0.81 |
| | Mortality | | | | | | | | AUC | 0.75 |
| | | | | | | | | | Sensitivity | 0.71 |
| | | | | | | | | | Specificity | 0.76 |
| Jordan<br>2022 (65) | Need for ICU | 2456<br>Train: 2000<br>Test: 456<br>F: *NA*<br>M: *NA* | 55.3 | CXR | Classification | P: rescaling images to an isotropic resolution, resampling and normalizing<br>A: *NA* | CNN | Geographical extent of airspace opacities | AUC | 0.87 |
| | Mortality | | | | | | | | | 0.82 |
| Bae<br>2021 (25) | Ventilation requirement | 691<br>(*NA*)<br>F:328<br>M:363 | 56 | CXR | Classification | P: Segmentation of lung and artifact, Average histogram matching, Automatic cropping<br>A: Flipping, Rotation, Translation | Machine learning (RF, LDA and QDA) | Image features | AUC | 0.78 |
| | | | | | | | | | Sensitivity | 0.72 |
| | | | | | | | | | Specificity | 0.72 |
| | | | | | | | CNN | | AUC | 0.75 |
| | | | | | | | | | Sensitivity | 0.64 |
| | | | | | | | | | Specificity | 0.73 |
| | Mortality | | | | | | Machine learning (RF, LDA and QDA) | | AUC | 0.78 |
| | | | | | | | | | Sensitivity | 0.70 |
| | | | | | | | | | Specificity | 0.73 |
| | | | | | | | CNN | | AUC | 0.75 |
| | | | | | | | | | Sensitivity | 0.59 |
| | | | | | | | | | Specificity | 0.74 |
| Aljouie 2021 (67) | Ventilation requirement | 1508 Train: 1208 Test: 300 F:651 M:857 | 54.8 | CXR | Classification | P: Feature normalization, Feature selection A: SMOTE, ADASYN sampling approach, RUS | BRF | Image features | AUC | 0.76 |
| | | | | | | | | | Accuracy | 0.52 |
| | Mortality | 1513 Train: 1212 Test: 301 F: 653 M:860 | 54.8 | | | | XGB + ADASYN | | AUC | 0.74 |
| | | | | | | | | | Accuracy | 0.71 |
| Bermejo 2022 (63) | Mortality | 103 (60/93) M=39 F=64 | 64.83 | CT | segmentation | P: clipping the intensities outside the range, rescaling A: NA | CNN | Lesion region | AUC | 0.87 |
| | Admission to the Intensive Care Units (ICU) | | | | | | | | | 0.73 |
| | Need for mechanical ventilation | | | | | | | | | 0.68 |
| Aslam 2022 (68) | Mortality | 1513 (136/1377) | NA | CXR | classification | NA | EAI* and DL Model | Image features | AUC | 0.998 |
| | | | | | | | | | Balanced accuracy | 0.98 |
| | Ventilator support | 1508 (295/1213) | | | | | | | AUC | 0.98 |
| | | | | | | | | | Balanced accuracy | 0.97 |
| Ahmedt 2022 (69) | Mortality | 673 (148/390) | NA | CXR | classification | P: + A: Horizontal flip, rotation, shear, and zoom | Xception, | Image features | F1 | 83.91 |
| | | | | | | | InceptionResNet V2 | | | 96.28 |
| | | | | | | | VGG 16 | | | 99.56 |
| | | | | | | | VGG 19 | | | 99.79 |
| Shiri 2022 (71) | Mortality | 14339 Train:10038 Test:4301 F:6722 M:7617 | *NA* | CT | Classification | P: cropped to the lung region and then resized to 296 × 216 ,image voxel was resized to an isotropic voxel size of 1 × 1 × 1 mm3 ,intensity discretized to 64-binning size A: *NA* | ANOVA feature selector + Random Forest (RF) classifier | Intensity and texture radiomics features | AUC | 0.83 |
| | | | | | | | | | Sensitivity | 0.81 |
| | | | | | | | | | Specificity | 0.72 |
| Spagnoli 2022 (72) | Mortality | 436 Test:436 F:150 M:286 | 68.5 | CT | Classification | P: Synthetic Minority Oversampling Techniques for random foreast model. Normalization and scaling of the features for LASSO and FcNN. A: *NA* | Fully connected Neural Network (FcNN) | Lung consolidation, ground glass opacity, crazy paving and bilateral involvement | AUC | 0.62 |
| | | | | | | | | | Sensitivity | 0.63 |
| | | | | | | | | | Specificity | 0.56 |
| | | | | | | | Least Absolute Shrinkage and Selection Operator (LASSO) | | AUC | 0.61 |
| | | | | | | | | | Sensitivity | 0.70 |
| | | | | | | | | | Specificity | 0.51 |
| Shiri 2021 (70) | Mortality | 152 Train: 106 Test: 46 F: 65 M:87 | 61.1 | CT | Classification | P: Interpolation to isotropic voxel, Re-sample to 1×1×1 $mm^3$ A: *NA* | MRMR and XGB | Image features | AUC | 0.91 |
| | | | | | | | | | Accuracy | 0.83 |
| | | | | | | | | | Sensitivity | 0.85 |
| | | | | | | | | | Specificity | 0.80 |

M: Male/ F: Female/ AUC: Area Under Curve/ OS1: 20 point CT score/ OS2: 25 point CT score/ OP: Opacity Persentage/ DSC: Dice Similarity Coefficient/ SVM: Support Vector Machine/ ED CNN: Encoder-Decoder Convolutional Neural Network/ FPN: Feature Pyramid Network/ NA: Not Available/ MGRF: Markov-Gibbs Random Field/ RF: Random Forest/ KNN: K-Nearest Neighbors/ TSCR: T Test Score/ BRF: Balanced Random Forest classifier/ XGB: eXtreme Gradient Boosting/ ADASYN: Adaptive Synthetic sampling approach/ MRMR: Maximum Relevance Minimum Redundancy/ MOI: Mass of infection/ POI: Persentage of infection/ LDA: Linear Discriminant Analysis/ QDA: Quadratic Discriminant Analusis/ CT: Computed Tomography/ CXR: Chest X-ray/ CDF: Cumulative Distribution Function

**Table 4**. Summary of the Meta-analysis Statistics

| Parameter | | Mortality | Severity Assessment | Need for Ventilation |
|---|---|---|---|---|
| **No. Studies** | | 4 | 9 | 4 |
| **No. Models** | | 7 | 13 | 6 |
| **Pooled Sensitivity** | | 71%; 95% CI [65%, 77%] | 88%; 95% CI [77%, 94%] | 67%; 95% CI [61%, 73%] |
| **Pooled Specificity** | | 69%; 95% CI [61%, 76%] | 89%; 95% CI [82%, 94%] | 89%; 95% CI [75%, 95%] |
| **Positive Likelihood Ratio** | | 2.3; 95% CI [1.7, 3.1] | 8.2; 95% CI [4.6, 14.5] | 5.9; 95% CI [2.7, 13.1] |
| **Negative Likelihood Ratio** | | 0.41; 95% CI [0.31, 0.55] | 0.14; 95% CI [0.07, 0.28] | 0.37; 95% CI [0.32, 0.43] |
| **Diagnostic Odds Ratio (DOR)** | | 6; 95% CI [3, 10] | 59; 95% CI [18, 197] | 16; 95% CI [7, 36] |
| **AUC (HSROC)** | | 0.76; 95% CI [0.72, 0.80] | 0.95; 95% CI [0.92, 0.96] | 0.77; 95% CI [0.73, 0.80] |
| **Heterogeneity (Chi-square)** | **Q** | 94.059 | 102.071 | 365.338 |
| | **df** | 2.00 | 2.00 | 2.00 |
| | **p-value** | **0.000** | **0.000** | **0.000** |
| **Inconsistency (I-square) – $I^2$** | | 98%; 95% CI [97%- 99%] | 98%; 95% CI [97%- 99%] | 99%; 95% CI [99%- 100%] |
| **Proportion of heterogeneity likely due to threshold effect** | | 0.13 | 0.51 | 0.49 |
| **Deek's Funnel Plot asymmetry test p-value** | | 0.19 | 0.07 | 0.92 |